\documentclass[aps,amsmath,amssymb,twocolumn,showpacs,prl]{revtex4-1}
\usepackage{graphicx,color}
\graphicspath{{./}{./figures/}}
\usepackage{dcolumn}% Align table columns on decimal point
\usepackage{bm}% bold math
\newcommand{\be}{\begin{equation}}
\newcommand{\ee}{\end{equation}}
\newcommand{\bea}{\begin{eqnarray}}
\newcommand{\eea}{\end{eqnarray}}

\newcommand{\rmd}{\mathrm{d}}

\newcommand{\mO}{\mathcal{O}}
\newcommand{\hG}{\widehat{G}}

\def\bea#1\eea{\begin{align}#1\end{align}}

\begin{document}
\title{Explore or exploit? A generic model and an exactly solvable case}
\author{Thomas Gueudr\'e, Alexander Dobrinevski, Jean-Philippe Bouchaud} \affiliation{CNRS-Laboratoire
de Physique Th{\'e}orique de l'Ecole Normale Sup{\'e}rieure, 24 rue
Lhomond, 75231 Cedex 05, Paris, France\\
Capital Fund Management, 21 rue de l'Universt\'e, 75007 Paris, France.} 
\date{\today\ -- \jobname}
%\date{\today\ -- \jobname\ -- compilation \input{\jobname.counter}}


\begin{abstract}
Finding a good compromise between the exploitation of known resources and the exploration of unknown, but potentially more profitable choices, is
a general problem, which arises in many different scientific disciplines. We propose a stylized model for these exploration-exploitation situations,
including population or economic growth, portfolio optimisation, evolutionary dynamics, or the problem of optimal pinning of vortices or dislocations in disordered materials. 
We find the exact growth rate of this model for tree-like geometries and prove the existence of an optimal migration rate in this case. Numerical simulations in 
the one-dimensional case confirm the generic existence of an optimum.
\end{abstract}

\maketitle
The exploration-exploitation tradeoff problem pervades a large number of different fields (see \cite{Gen} and the many references therein). 
Two early examples concern the management of firms \cite{Exp1} (should one exploit 
an already known technology or explore other avenues, potentially more profitable, but risky?) and the so-called multi-arm bandit problem \cite{Exp3} 
(sticking with the 
seemingly most profitable arm to date, or switching in search of potentially more profitable ones?). 
Clearly, this is a universal paradigm that ranges from population growth and animal foraging to economic growth, investment strategies or optimal research 
policies. As we will show below, the same issues also arise, in a slightly disguised form, in the context of vortex or dislocation pinning by impurities, 
and are relevant for material design. Intuitively, neither staying at the same place (and missing interesting opportunities) nor 
changing places too rapidly (and failing to exploit favorable circumstances) are optimal strategies. An optimal, non zero search rate should thus exist in general.
However, there are no exactly solvable cases where the exploration-exploitation tradeoff can be investigated in details.  
The aim of this paper is to propose a general, stylized model for these exploration-exploitation situations, which encompasses all the examples given above. 
We obtain exact solutions of this model in two cases (a fully connected and a tree geometry), for which we explicitely prove the existence of a non-trivial
optimal search rate. Euclidean geometries are also considered, as these correspond to physical situations, like the pinning problem alluded to above. In this
case, perturbation theory and numerical simulations confirm the existence of an optimum as well. 

Our model describes the dynamics of a quantity we generically call $Z_i$, defined on the nodes $i$ of an arbitrary graph, that evolves according to
the following equation\cite{BM}:
\be\label{Eq-gen}
\frac{\partial Z_i(t)}{\partial t} = \sum_{j \neq i} J_{ij} Z_j(t) - \sum_{j \neq i} J_{ji} Z_i(t) + \eta_i(t) Z_i(t).
\ee
The first two terms encode ``migration'' effects, with $J_{ij}$ the migration rate from $j$ to $i$. The last term describes the growth (or decay) of the
quantity $Z_i$ with a random growth rate $\eta_i(t)$. We will choose $\eta_i$ to be Gaussian, centred and uncorrelated from site to site, with a exponential time-correlator:
\be
\label{eq:CorrOU}
\langle \eta_i(t_1) \eta_j(t_2) \rangle = \delta_{ij} \frac{\sigma^2}{2 \tau} e^{-\frac{|t_1-t_2|}{\tau}}.
\ee
Our qualitative conclusions are however independent of the precise form \eqref{eq:CorrOU}, provided correlations decay on a {\it finite scale} $\tau$, 
which will play an important role in the following. 

Many different problems are described by Eq. (\ref{Eq-gen}). Population dynamics (bacteria, humans, animals) is one example with $Z_i$ 
the number of individuals around site (or habitat) $i$. In this setting, $\eta_i(t)$ encodes the local balance between beneficial and detrimental effects on population growth \cite{Nelson} 
(i.e. quality and quantity of resources/nutrients, climate, illnesses, etc.).
A slightly different interpretation can be given in the context of evolutionary dynamics, where the sites $i$ correspond to different alleles 
and the $J_{ij}$ are mutation rates. 
In the context of pinning problems, $Z_i$ corresponds to the partition function of a linear object of length $t$ (polymers, vortices, dislocations), 
ending on site $i$, that can
hop between sites and interact with a local random pinning potential $\eta_i(t)$ \cite{HHZ}. In an economics setting, Eq.(\ref{Eq-gen}) can be interpreted as describing the dynamics of the
wealth of individuals that exchange and invest in risky projects, or of the total activity in a sector of the economy $i$, that may shift from one sector to another, and grow or 
decay depending on innovations, raw material prices, etc. Another interesting application is that of portfolio theory, where $Z_i$ is the amount of money invested in asset $i$ \cite{Cover}.
Then $\eta_i(t)$ is the return streams of this assets and the $J_{ij}$ describe the reallocation of the gains made on some assets towards the rest of the portfolio.
Without this rebalancing the portfolio would end up being concentrated in one (or a few) assets only (see e.g. \cite{BP}, pp. 37-38), and hence be exceedingly risky. 

In the case where $J_{ij} \equiv J$ and the nodes $i$ are on a regular lattice in $d$ dimensions, Eq. (\ref{Eq-gen}) 
is a discretized version of the ``stochastic heat equation'', 
\be\label{Eq-stoheat}
\frac{\partial Z(\vec x,t)}{\partial t} = J \nabla^2  Z(\vec x,t) + \eta(\vec x,t)  Z(\vec x,t).
\ee
Upon a Cole-Hopf transformation $Z = e^h$, this equation maps into the celebrated KPZ equation $\partial_t h = J \nabla^2 h + J (\nabla h)^2 + \eta$ 
that appears in a wide variety 
of domains: cosmology \& turbulence \cite{Bec,BecKhanin}, surface growth \cite{Krug,Barabasi}, directed polymers \cite{HHZ} or Hamilton-Jacobi-Bellmann optimisation problems
\cite{Kappen}. 

A host of exact results have recently been obtained for the one dimensional ($d=1$) case, in particular concerning the scaling properties 
of the {\it fluctuations} of the $h$-field (for a review, see \cite{Corwin}). 
Here, however, we will not be concerned with these fluctuations but interested in the long-time average ``velocity'' $c$ of the $h$-field, 
defined in the discrete case as:
\be
c := \lim_{t \to \infty} \frac{1}{Nt} \sum_{i=1}^N \ln Z_i(t),
\ee
where $N$ is the total number of sites. This velocity $c$ has a clear interpretation in all the examples mentioned above: it represents the average asymptotic 
growth rate of the population, or of the economic wealth in models of growth, the free-energy of the polymer, vortex, etc. in the context of pinning. 
It is therefore very 
natural to look for the maximum of this quantity as a function of the parameters of the model, since these will correspond to optimal situation -- 
either in terms of population, economic or portfolio growth, or in terms of pinning efficiency, which is relevant for material design, 
for example superconductors with high critical currents \cite{HTC}. In this case, so-called ``columnar disorder'' \cite{KrugHH} (corresponding to a time correlated random noise 
$\eta$ in the present language) is known to be highly effective at pinning vortices \cite{PLD-Nelson,GiamarchiLedoussal}. 
Our central result is that for non-zero correlation
time of the random noise/potential $\eta_i(t)$, there exists an optimal migration rate $J$ such that $c$ reaches a maximum. 
This optimal rate realizes the ``exploration-exploitation'' 
compromise: moving too slowly ($J$ small) does not allow the system to probe the environment efficiently, and some favorable opportunities are missed. 
Moving too fast ($J$ large), on the other hand, 
does not allow the system to fully benefit from favorable spots that last for a time $\sim \tau$, as it leaves these spots too early.

Let us first present numerical simulations of the 1+1 directed polymer problem with time-correlated disorder. The equation we simulated is:
\begin{align}
\nonumber
Z_i(t+\rmd t) &=(1-2 J \rmd t) Z_i(t)+ \\
J \rmd t&\left[Z_{i+1}(t)+Z_{i-1}(t)\right]+ \eta_i(t)\rmd t Z_i(t),
\label{discreteSHE}
\end{align} 
with $i=1, \dots, N$ and $\eta_i(t)$ an exponentially correlated Gaussian noise, as in Eq. (\ref{eq:CorrOU}). 
We considered a system with $N=512$ sites and periodic boundary conditions. We determined $c(J)$ after a time $t=40$ 
long enough to reach a stationary state, and much greater than the correlation time fixed here to $\tau=0.1$. The dependence of $c$ on $J$ for $\sigma=1$ and $\tau=0.1$ 
is shown in Fig. 1, together with a) the result of direct perturbation theory of the KPZ equation, a priori valid for large $J$, and 
b) the prediction of the ``tree-approximation" with $a=1/2$ and $m=1$ that we detail
below. The former predicts  $c(J)\approx \sigma^2/4\sqrt{J\tau} + O(\sigma^4/J)$ for $J \to \infty$, which indeed 
fits the data quite well in the large $J$ region, without any adjustable parameter. The tree-approximation, on the other hand, is quantitatively incorrect as expected for
a one-dimensional system. For example, it predicts a $J^{-1}$ decay of $c$ (see below) but still manages to capture approximately the overall behaviour of $c(J)$, in particular 
the existence of a maximum. 

\begin{figure}%
\includegraphics[width=\columnwidth]{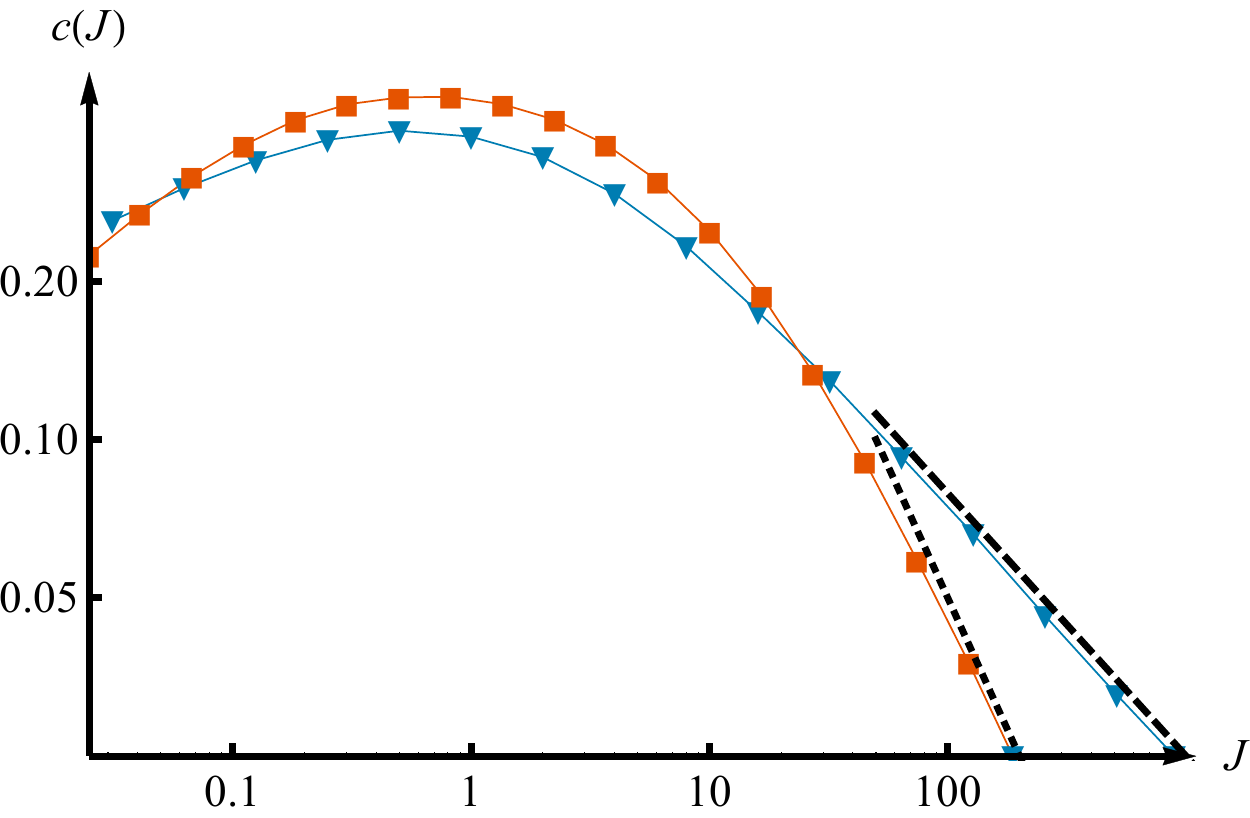}%
\caption{Comparison of simulations of \eqref{discreteSHE} (blue triangles) and of \eqref{eq:ModelRandomBranch} (red squares) 
for $N=512$ and $\tau=0.1$, as functions of the branching (diffusion) rate {$J$}. 
We fix $\sigma=1$. The dashed black line is the large {$J$} 
asymptotics obtained from perturbation theory, $c(J)\approx \sigma^2/4\sqrt{J\tau}$. The dotted line reveals the $J^{-1}$ behaviour of $c(J)$ 
for large $J$, predicted by the tree-approximation (with $a=1/2$ and $m=1$).}%
\label{fig:PlotVariousTau}%
\end{figure}

Let us now turn to a simplified model, where the interplay between exploration and exploitation, and the optimal migration rate, can be fully understood analytically. 
We first note that our general model Eq. (\ref{Eq-gen}) for a regular lattice with $J_{ij}=J$ for neighbouring sites, can be slightly altered as the following evolution rule:
\bea
\label{eq:ModelRandomBranch}
Z_i(t+\rmd t) = 
\begin{cases}
Z_i(t+\rmd t) \exp\left[\eta_i(t)\rmd t \right]& \text{prob.}\, 1-J \rmd t, \\
(1 - {a}) Z_i(t) + \frac{a}{m} \sum_{j \wedge i}  Z_j(t)&  \text{prob.}\, J \rmd t.
\end{cases}
\eea
where $m$ is the number of neighbours of $i$ and $j \wedge i$ means that $i,j$ are neighbours. 
To obtain a solvable model, we neglect all spatial correlations between the $Z_i$'s, 
which amounts to the {\it tree approximation} introduced by Derrida and Spohn for the
directed polymer problem in 1988 \cite{DerridaSpohn1988}. Following these authors, we define the generating functions
\bea
\nonumber
G_t(x,\eta) :=& \left\langle \exp\left[-e^{-x}Z_i(t)\right]\delta\left[\eta_i(t)-\eta\right] \right\rangle, \\
\label{eq:RandomBranchGenFct}
\hG_t(x) :=& \int_{-\infty}^\infty \rmd \eta\, G_t(x,\eta) = \left\langle \exp\left[-e^{-x}Z_i(t)\right]\right\rangle
\eea
Assuming the $Z_i$'s to be independent allows one to write the following evolution equation for $G_t(x,\eta)$:
\bea
\nonumber
& G_{t+\rmd t}(x,\eta) =    \\
\nonumber
&(1-J\, \rmd t)\langle \exp\left[-e^{-x+\eta_i(t)\rmd t}Z_i(t)\right]
\delta\left[\eta_i(t+\rmd t) -\eta\right] \rangle   \\
\nonumber
\label{eq:EvolG1}
& + J \, \rmd t\left\langle \exp\left[-e^{-x+q_1}Z_i(t)\right]\delta\left[\eta_i(t)-\eta\right]\right\rangle   \\
& \times \left\langle \exp\left[-e^{-x+q_2}Z_j(t)\right]\right\rangle ^m
\eea
with  $q_1 := \log(1 - {a})$ and $q_2 := \log \frac{a}{m}$. 
The choice in Eq.~\eqref{eq:CorrOU} of an Ornstein-Uhlenbeck process for $\eta$ is particularly simple, 
since it yields a \textit{Markovian} equation for $\eta_i(t+\rmd t)$:
\bea
\eta_i(t+\rmd t) = \eta_i(t) + \left[ - \frac{1}{\tau_c}\eta_i(t) + \frac{\sigma}{\tau_c} \xi_i(t)\right] \rmd t,
\eea
where $\xi$ is a Gaussian white noise. Inserting this into \eqref{eq:EvolG1}, and expanding to $\mO(\rmd t)$, we obtain
\bea
\nonumber
G_{t+\rmd t}(x,\eta) =& \left\langle G_t\left(x, \eta + \frac{\sigma}{\tau_c}\xi_i(t) \rmd t \right) \right\rangle_{\xi} - J\, \rmd t G_t(x,\eta) \\
\nonumber
& - \eta\, \rmd t\, \partial_x G_t(x,\eta) + \frac{\rmd t}{\tau} \partial_\eta\left[\eta G_t(x,\eta)\right]  \\
\nonumber
&+ J \, \rmd t\, G_t(x - q_1,\eta) \hG_t(x-q_2)^m + \mathcal{O}(\rmd t)^2
\eea

Using $\xi_i(t)\rmd t \sim \mathcal{N}(0,\rmd t)$, we now average over $\xi$ and 
obtain a generalized Fisher-KPP equation for $G$, where the diffusion operator is replaced by the Ornstein-Uhlenbeck 
operator, involving the additional state variable $\eta$:
\bea
\nonumber
\partial_t G_t(x,\eta) = & \frac{\sigma^2}{2\tau_c^2}\partial_\eta^2 G + \frac{1}{\tau_c} \partial_\eta \left(\eta G \right) - \eta \partial_x G  \\
\label{eq:RandomBranchPDE}
& + J \left[G_t(x - q_1,\eta) \hG_t(x-q_2)^m - G_t(x,\eta)\right].
\eea
Like the Fisher-KPP equation known from the standard mean-field directed polymer problem \cite{DerridaSpohn1988,CookDerrida1990}, 
it gives rise to a front propagating in the $x$ direction. The velocity of this front is precisely the quantity $c$ we are looking for 
and is fixed by the tail behaviour of 
$G_t$ when $x \to \infty$.  In this tail, we make the following ansatz for $G$:
\be
\label{eq:RandomBranchGTailEta}
G_t(x,\eta) = Q(\eta) - R(\eta)e^{-\gamma(x-ct)} + ...
\ee
with $\int \rmd \eta\, Q(\eta)=1$. Inserting this into \eqref{eq:RandomBranchPDE}, one finds, by identifying terms of order $1$ and 
terms of order $e^{-\gamma(x-ct)}$, that $Q(\eta)$ is the stationary Gaussian distribution for the Ornstein-Uhlenbeck process $\eta(t)$ (as it should be), 
while $R(\eta)$ satisfies:
\bea
\nonumber
R c \gamma = & \frac{\sigma^2}{2 \tau^2} \partial^2_\eta R + \frac{1}{\tau} \partial_\eta (\eta R) + \eta R + J m e^{\gamma q_2} Q \int \rmd\eta R(\eta) \\
\label{eq:RandomBranchR}
&  + R(\eta) J (e^{\gamma q_1}-1).
\eea
This can be simplified by imposing (without loss of generality) $\int \rmd \eta\, R(\eta)=1$ and setting $R = \phi e^{-\eta^2 \tau/2\sigma^2}$, 
$\sigma^2\hat{c} = \gamma c - J(e^{\gamma q_1}-1)-\sigma^2\gamma^2/2$ and $y = \eta/\sigma^2 - \gamma$. One gets the following equation for $\phi$:
\be
- \frac{1}{2 \sigma^4 \tau^2} \phi'' + \frac{1}{2} y^2 \phi + (\hat c -\frac{1}{2 \sigma ^2 \tau}) \phi = 
\frac{Jm e^{\gamma q_2}}{\sigma^2} \frac{e^{-\frac{(y+1)^2 \sigma^2 \tau}{2}}}{\sqrt{\pi \sigma^2/\tau}}.
\ee
Introducing the harmonic oscillator eigenfunctions $\phi_n(y) =  e^{- y^2 \sigma^2 \tau /2} H_n(y \sigma \sqrt{\tau})$, the 
solution of the above equation can be written as $\phi(y) = \sum_{n=0}^\infty A_n \phi_n(y)$ where the coefficients $A_n$ 
are given by:
\bea
A_n \left[\frac{n}{\sigma^2 \tau} + \hat c \right] = \frac{Jm e^{\gamma q_2}}{\sigma^2} 
\frac{e^{- \gamma^2 \sigma^2 \tau/4}\frac{(-1)^n}{n!} \left(\frac{\sigma\sqrt{\tau}}{2}\right)^n}{\sqrt{\pi \sigma^2/\tau}}.
\eea
Finally, the condition $\int \rmd \eta\, R(\eta)=1$ yields an implicit equation for $c$, valid for arbitrary $\tau$ 
\footnote{The infinite sum over $n$ can be rewritten in a integral form that is convenient for an asymptotic analysis of the equation.}
%\footnote{
%The infinite sum over $n$ can be rewritten as:
%\be
%\int_0^{\infty} {\rm d} s\,\exp\left\{-\left[\gamma c - J\left(e^{\gamma q_1}-1\right) - \frac{\sigma^2}{2}\right]s  
%+ \frac{\gamma^2 \sigma^2 \tau}{2} e^{-s/\tau}\right\},
%\ee
%which is convenient for an asymptotic analysis of the equation.
%}
:
\be
\label{eq:RandomBranchSpectrum}
1 = Jm e^{\gamma q_2- \gamma^2 \sigma^2 \tau/2}  \sum_{n=0}^\infty \frac{(\gamma^2 \sigma^2 \tau/2)^n}{n! \left[\frac{n}{\tau} + \gamma c -J(e^{\gamma q_1} - 1) - \frac{\sigma ^2}{2}\gamma^2\right]}
\ee
As in the Derrida-Spohn case, the corresponding function $c(\gamma)$ is found to reach a minimum value for a certain $\gamma_{\min}$, that depends on the parameters $J, \sigma^2, \tau, m$.
The interpretation of this phenomenon is now standard: only traveling waves with $\gamma \leq \gamma_{\min}$ can be sustained, and propagate at the speed $c(\gamma)$. 
A wave front which is ``too sharp'', i.e. prepared initially with a $\gamma > \gamma_{\min}$, will broaden until it reaches $\gamma = \gamma_{\min}$, and will propagate 
with the velocity $c(\gamma_{\min})$. In our case, the initial condition $Z=1$ corresponds to $\gamma = 1$; therefore either $\gamma_{\min}$ is found to be larger than unity, 
in which case $c$ is given by the solution of Eq. \eqref{eq:RandomBranchSpectrum} with $\gamma = 1$, 
or $\gamma_{\min} \leq 1$, in which case $c = c(\gamma_{\min})$. For the directed polymer/pinning problem, the first case corresponds to  the high-temperature, annealed phase 
(arising for $J \geq J_c$), while the second case corresponds to a low-temperature, frozen phase (for $J \leq J_c$). In the random growth problems, the latter case corresponds to a localization 
of the population/wealth/portfolio on a small number of particularly favorable habitats/individuals/assets (see the discussion in \cite{BM}). 

We determine $c(\gamma)$, $\gamma_{\min}$ and $c(\gamma_{\min})$ numerically from \eqref{eq:RandomBranchSpectrum}, 
with very good agreement with numerical simulations (see figure \ref{fig:PlotVariousTau}). We see in particular that for $\tau=0$, 
increasing the migration rate always {\it increases} the growth rate, which saturates at a constant value $c=\sigma^2/2$, 
for all $J \geq J_c$. Therefore, no optimum tradeoff between exploration and exploitation exists in this case -- exploring is always favorable or neutral. 
However, when a finite correlation time 
$\tau$ is introduced, we see that, as expected, an optimum migration rate indeed appears 
(cf. figure \ref{fig:PlotVariousTau}).\footnote{Note that the ``columnar'' limit where $\tau \rightarrow \infty$ \cite{KrugHH,PLD-Nelson,GiamarchiLedoussal}, 
is not easily approachable through the KPP mapping due to the non commutativity of the large $t$ and large $\tau$ limits.} 
In particular, we find analytically that for small $J$, $c(J) = \sqrt{2 \sigma ^2 J} + \mO(J)$ while for large $J$, $c(J) = \sigma^2/{2 a J \tau} + \mO(J)^{-2}$. 
In fact, the large $J$ behaviour can be understood heuristically as follows. Clearly, the problem for $\tau > 0$ must be equivalent, for large times, to the
standard uncorrelated case ($\tau=0$), but with a renormalized disorder amplitude. For large $J$ and finite $\tau$, the 
disorder cannot change the random walk nature of the exploration up to time $\tau$. The walk therefore freely visits $N_{\neq}=\mO(J \tau)$ 
different sites during this time, leading to a pre-averaging of the random disorder that reduces the variance $\sigma^2$ by a 
factor $N_{\neq}$. Since for $\tau=0$, $c \propto \sigma^2$, the above renormalisation immediately leads to $c(J) \sim \sigma^2/{J \tau}$ 
at large $J$. [Note that the very same argument leads to $c(J) \sim \sigma^2/\sqrt{J \tau}$ in $d=1$, as found above, and is also exact in $d=2$, where
logarithmic corrections appear.] Now since  $c(J=0)=0$ trivially, the decaying behaviour of $c(J)$ for large $J$ and finite $\tau$ immediately implies the generic existence of an 
optimum in the exploration rate, as anticipated above.

\begin{figure}%
\includegraphics[width=\columnwidth]{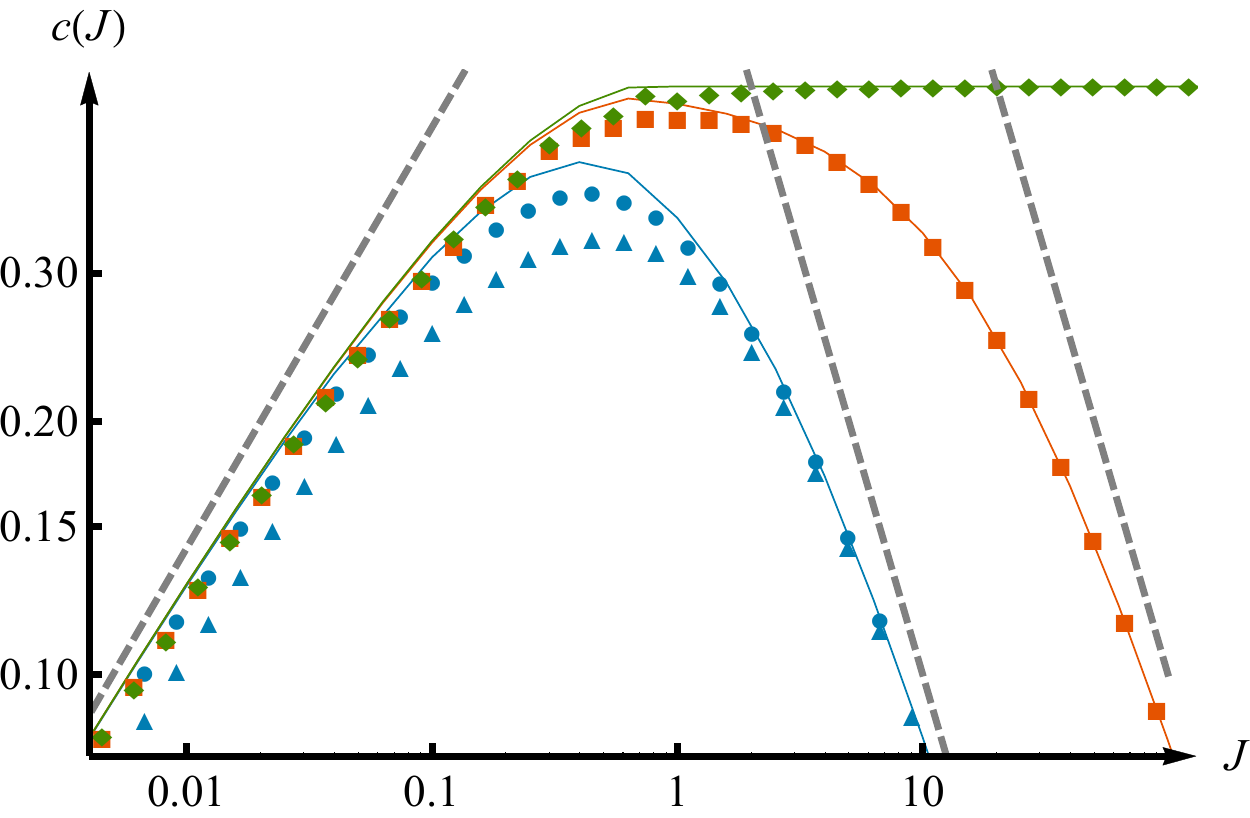}%
\caption{(Color online) Comparison of simulations of \eqref{eq:ModelRandomBranch} 
for various $N$ and $\tau$, as a function of the branching (diffusion) rate {$J$}. Green diamonds, orange squares and blue circles were obtained 
from numerical simulation of \eqref{eq:ModelRandomBranch} for $\tau=0, 0.1$ and $1$ with $N=2^{20}$. Blue triangles correspond to $\tau=1$ and $N=2^8$.
In all cases, $\sigma=1$ and $a=1/2$, $m=1$. The solid curves were obtained by numerical solution 
of \eqref{eq:RandomBranchSpectrum} for the corresponding values of $\tau$. The dashed grey lines are the large and small-{$J$} 
asymptotics. }%
\label{fig:PlotVariousTau}%
\end{figure}

We find very similar conclusions \cite{isunpublished} for another exactly solvable limit, the fully connected graph where $J_{ij}=J_0/N$, $\forall i,j$, 
which in fact corresponds (up to minor details) to the limit $a=1-\frac1N$ and $m=N$ of the tree model above.  Other theoretical methods used to investigate the KPZ/Directed Polymer
problem could also be useful to characterize $c(J)$ in $d+1$ dimensions or for other geometries, such as Mode-Coupling Theory or 
the Gaussian Variational method. The mapping to interacting bosons in the 1+1 case is also an interesting avenue we are exploring 
\cite{isunpublished}. It would be very interesting to observe the predicted pinning optimum experimentally. One possibility is in superconductors where the hopping
rate $J$ is related to the elastic energy of the vortex lattice, which itself depends on the external magnetic field. Changing the temperature 
is also a way to affect both the hopping constant and the effective pinning strength \cite{RossoPLD}. Applications of these ideas are numerous, in 
particular to quantify how diversified portfolios benefit from a balance between persistence and rebalancing, or to understand 
how economic growth is impacted by the ability of societies to find a tradeoff between tradition and innovation, or else collapse \cite{collapse}.

We thank G. Biroli, P. Le Doussal and R. Munos for very helpful insights.

\bibliographystyle{unsrt}
\bibliography{migration}

\begin{thebibliography}{10}

\bibitem{Gen}
J.~D. Cohen, S.~M. {McClure}, and A.~J. Yu.
\newblock {\em Philos Trans R Soc Lond B Biol Sci}, 362(1481):933--942, 2007.

\bibitem{Exp1}
J.~G. March.
\newblock Exploration and exploitation in organizational learning.
\newblock {\em Organization Science}, 2(1):71--87, 1991.
\newblock Note that it has 11,000 citations at the time of writing !

\bibitem{Exp3}
J.~C. Gittins and D.~M. Jones.
\newblock A dynamic allocation index for the sequential design of experiments.
\newblock In {\em Progress in statistics (European Meeting Statisticians,
  Budapest, 1972)}, pages 241--266. North-Holland, Amsterdam, 1974.

\bibitem{BM}
J.P. Bouchaud and M.~M\'ezard.
\newblock Wealth condensation in a simple model of economy.
\newblock {\em Physica A}, 282(3–4):536--545, 2000.

\bibitem{Nelson}
D.~R. Nelson and N.~M. Shnerb.
\newblock Non-hermitian localization and population biology.
\newblock {\em Phys. Rev. E}, 58(2):1383--1403, 1998.

\bibitem{HHZ}
T.~Halpin-Healy and Y.-C. Zhang.
\newblock Kinetic roughening phenomena, stochastic growth, directed polymers
  and all that. aspects of multidisciplinary statistical mechanics.
\newblock {\em Physics Reports}, 254(4–6):215--414, 1995.

\bibitem{Cover}
T.~M. Cover.
\newblock Universal portfolios.
\newblock {\em Mathematical Finance}, 1:1--29, 1991.

\bibitem{BP}
J.-P. Bouchaud and M.~Potters.
\newblock {\em Theory of Financial Risks and Derivative Pricing}.
\newblock Cambridge University Press, 2003.

\bibitem{Bec}
U.~Frisch and J.~Bec.
\newblock Burgulence.
\newblock In M.~Lesieur, A.~Yaglom, and F.~David, editors, {\em New trends in
  turbulence Turbulence: nouveaux aspects}, pages 341--383. Springer Berlin
  Heidelberg, 2001.

\bibitem{BecKhanin}
J.~Bec and K.~Khanin.
\newblock Burgers turbulence.
\newblock {\em Physics Reports}, 447(1–2):1 -- 66, 2007.

\bibitem{Krug}
J.~Krug and H.~Spohn.
\newblock {\em Kinetic roughening of growing surfaces in Solids Far from
  Equilibrium}.
\newblock C. Godr\`eche, Cambridge University Press, 1991.

\bibitem{Barabasi}
A.-L. Barabasi and H.~E. Stanley.
\newblock {\em Fractal Concepts in Surface Growth}.
\newblock Cambridge University Press, 1995.

\bibitem{Kappen}
H.~J. Kappen.
\newblock Linear theory for control of nonlinear stochastic systems.
\newblock {\em Phys. Rev. Lett.}, 95(20):200--201, 2005.

\bibitem{Corwin}
I.~Corwin.
\newblock The {K}ardar-{P}arisi-{Z}hang equation and universality class.
\newblock {\em Random Matrices: Theory and Applications}, 01(01):1130001, 2012.

\bibitem{HTC}
B.~Maiorov and al.
\newblock Synergetic combination of different types of defect to optimize
  pinning landscape using {BaZrO3-doped} {YBa2Cu3O7}.
\newblock {\em Nat Mater}, 8(5):398--404, 2009.

\bibitem{KrugHH}
J.~Krug and T.~Halpin-Healy.
\newblock Directed polymers in the presence of columnar disorder.
\newblock {\em J. Phys. I France}, 3(11), 1993.

\bibitem{PLD-Nelson}
T.~Hwa, P.~Le~Doussal, D.~Nelson, and V.~Vinokur.
\newblock Flux pinning and forced vortex entanglement by splayed columnar
  defects.
\newblock {\em Phys. Rev. Lett.}, 71(21):3545--3548, 1993.

\bibitem{GiamarchiLedoussal}
T.~Giamarchi and P.~Le~Doussal.
\newblock Variational theory of elastic manifolds with correlated disorder and
  localization of interacting quantum particles.
\newblock {\em Phys. Rev. B}, 53:15206--15225, 1996.

\bibitem{DerridaSpohn1988}
B.~Derrida and H.~Spohn.
\newblock Polymers on disordered trees, spin glasses, and traveling waves.
\newblock {\em J Stat Phys}, 51(5-6):817--840, 1988.

\bibitem{CookDerrida1990}
J.~Cook and B.~Derrida.
\newblock Polymers on disordered hierarchical lattices: A nonlinear combination
  of random variables.
\newblock {\em Journal of Statistical Physics}, 57(1-2):89--139, 1989.

\bibitem{Note1}
The infinite sum over $n$ can be rewritten in a integral form that is
  convenient for an asymptotic analysis of the equation.

\bibitem{Note2}
Note that the ``columnar'' limit where $\tau \rightarrow \infty $ \cite
  {KrugHH,PLD-Nelson,GiamarchiLedoussal}, is not easily approachable through
  the KPP mapping due to the non commutativity of the large $t$ and large $\tau
  $ limits.

\bibitem{isunpublished}
T.~Gueudr\'{e}, A.~Dobrinevski, and J-P. Bouchaud.
\newblock in preparation.

\bibitem{RossoPLD}
S.~Bustingorry, P.~Le~Doussal, and A.~Rosso.
\newblock Universal high-temperature regime of pinned elastic objects.
\newblock {\em Phys. Rev. B}, 82(14):140201, 2010.

\bibitem{collapse}
J.~M. Diamond.
\newblock {\em Collapse: How Societies Choose to Fail Or Succeed}.
\newblock Penguin, 2006.

\end{thebibliography}

\end{document}